\documentclass[prb,amsmath,amssymb,twocolumn,notitlepage]{revtex4-2}

\usepackage{times}
\usepackage[T1]{fontenc}
\usepackage{bm}
\usepackage{bbm}
\usepackage{float}
\usepackage{mathrsfs}
\usepackage[utf8]{inputenc}
\usepackage{pdfsync}
\usepackage[colorlinks=true,bookmarks=true,pdfborder={0 0 0},linkcolor=blue,urlcolor=blue,breaklinks=true,bookmarksnumbered=true]{hyperref}
\def \x{\bm{x}}

\def \s{\bm{s}}

\def \i{\text{\bf{i}}}
\def \f{\text{\bf{f}}}

\def \hs{\hat{s}}

\def \hu{\hat{u}}
\def \hO{\hat{O}}
\def \hS{\hat{S}}

\def \hB{\hat{B}}
\def \heta{\hat{\eta}}
\def \hD{\hat{D}}
\def \hI{\hat{I}}

\def \mT{\mathcal{T}}
\def \mH{\mathcal{H}}

\def \hilbert{\mathscr{H}}
\def \pt{\partial}

\def \sgn{\text{sgn}}

\def \tr{\text{Tr}}
\def \Xs0{X^{\sigma 0}}
\def \X0s{X^{0 \sigma}}

\newcommand{\abs}[1]{\lvert#1\rvert}

\newcommand{\ev}[1]{\mbox{$\langle #1 \rangle$}}
\newcommand{\dev}[1]{\mbox{$\langle\langle #1 \rangle\rangle$}}

\newcommand{\ket}[1]{\mbox{$| #1 \rangle$}}

\newcommand{\ua}{\uparrow}
\newcommand{\da}{\downarrow}

\newcommand{\eqdisp}[1]{Eq.~(\ref{#1})}
\begin{abstract} Recently, an {\it algebraic-dynamical theory} (ADT) for strongly interacting many-body quantum Hamiltonians in W. Ding, arXiv: 2202.12082 (2022). By introducing the complete operator basis set, ADT proposes a generic framework for systematically constructing dynamical theories for interacting quantum Hamiltonians, using quantum entanglement as the organizing principle. In this work, we study exact ADT solutions of interacting two-spin problems which can be used as "free theories" for perturbation study on relevant solvable limits. Then we perform ADT  perturbation calculations and obtain the correct ground state under the perturbation, which shows that the ADT framework is capable of constructing correct dynamical perturbation theories for strongly interacting quantum spin models. We also discuss the implication for relevant lattice models.  \end{abstract}
\date{\textit{<2021-03-16 Tue>}}
\title{Algebraic-Dynamical Theory for Quantum Spin-1/2: the Two-spin limit and Implications for Lattice Models}
\hypersetup{
 pdfauthor={Wenxin Ding\(^{1*}\), Chuanru Dai\(^{1}\), Zhengfei Hu\(^{1}\)},
 pdftitle={Algebraic-Dynamical Theory for Quantum Spin-1/2: the Two-spin limit and Implications for Lattice Models},
 pdfkeywords={},
 pdfsubject={},
 pdfcreator={wxding@ahu.edu.cn},
 pdflang={English}}
\begin{document}

\title{Algebraic-Dynamical Theory for Quantum Spin-1/2: the Two-spin limit and Implications for Lattice Models} \author{Wenxin Ding\(^{1*}\), Chuanru Dai\(^{1}\), Zhengfei Hu\(^{1}\)}
 \date{\today} \affiliation{$^1$School of Physics and Optoelectronics Engineering, Anhui University, Hefei, Anhui Province, 230601, China}
  \email{wxding@ahu.edu.cn}
 \maketitle

\section{Introduction}
\label{sec:orgd03efe1}

Quantum spin systems are one of the most important class of systems for condensed matter physics and statistical mechanics. They are the foundation for microscopic understanding of magnetic properties of materials, but also interplay with other degrees of freedom and contribute to affect other physical properties significantly. For example, magnetic orders and its fluctuations are widely believed to be an important ingredient in unconventional and high-\(T_c\) superconductivity.

Quantum spin systems can show very rich and different behaviors, depending on the dimensions, lattices,  interactions and the spin number \(S\). Theoretically, many quantum spins models serve as paradigms for physical phenomena and theories, such as the Ising models\cite{Ising1925} for phase transitions, Bethe ansatz\cite{bethe-1931-zur-theor-metal} of one-dimensional (1D) quantum spin chains for integrability and quantum dynamics, the Haldane chain\cite{haldane-1983-nonlin-field} and the Affleck-Lieb-Kennedy-Tasaki (AKLT) model\cite{affleck-1987-rigor-resul} for topological phases, etc.. Besides such solvable models, even richer phenomena are found in quantum spin models and materials, from the bosonic-like spin-wave or magnon excitations of magnetic systems, to the Luttinger liquid behavior\cite{tomonaga-1950-remar-bloch,luttinger-1963-exact-solub} in 1D quantum spin systems\cite{haldane-1994-demon-of}, and the exotic behaviors of quantum spin liquids\cite{Savary2017,Zhou2017} such as the Kitaev model\cite{kitaev-2006-anyon-exact}.

However, solutions for perturbation on solvable models remain difficult, not to mention the more realistic models. The main difficulty of quantum spin problems is that the quantum spin operators are \emph{noncanonical} bosons: their commutators are not \(\delta\)-functions.
For example, for the quantum spin-1/2 systems, while spin-1/2 operators of different lattice sites commute, operators on the same site satisfy both a set of anti-commutation relations and the \(SU(2)\) Lie algebra  simultaneously. The anti-commutation relation is sometimes termed the hard-core condition, hence the quantum spin-1/2s are also considered hard-core bosons. Therefore, problems of quantum spin-1/2 are strongly interacting in nature. Furthermore, such noncanonical operator algebras also invalidate Wick's theorem, which is the very foundation of standard Feynman diagram techniques.

Conventional approaches to quantum spin-1/2 problems avoid the noncanonicality by expressing the physical operators in terms of auxiliary \emph{canonical} operators, either bosonic or fermionic, such as Dyson-Maleev bosons\cite{Dyson1956,Maleev1958}, Schwinger bosons\cite{auerbach-2008-schwinger-boson}, Abrikosov pseudo-fermions or Schwinger fermions\cite{abrikosov-1965-elect-scatt,affleck-1988-large-nlimit} and Holstein-Primakoff bosons\cite{Holstein1940}, which often enlarges the Hilbert space. Such parton or slave-particle constructions were studied intensively previously. However, they do have strong limitations. Typically, one parton construction can be more convenient for certain types of problems but not so convenient for others, depending on the nature of the underlying ground state. For example, bosonic constructions are used more often for magnetically ordered systems. For frustrated systems with a paramagnetic ground state, either distinguishing or unifying states given by different constructions are difficult\cite{lu-2017-unific-boson}.
Furthermore, the constraints due to Hilbert-space enlargement make it difficult to develop dynamical perturbation theories on top of existing parton mean field theories.

More exact theoretical investigation has been scarce. Vaks, Larkin\cite{Vaks-1968-spin-waves} initiated  efforts of pursuing an perturbative formalism for Ising spins and Kondo\cite{Kondo1972} for quantum spin-1/2, both realizing and utilizing the fundamental rigorousness and controllability of the equations-of-motion approach. Both approaches treat the interacting vertex functions approximately with a decoupling scheme which is termed ``random phase approximation''(RPA) or Tyablikov-decoupling \cite{Kondo1972,tyablikov-1967-method-quant,frobrich-2006-many-body}. In strong coupling problems, such decoupling is not justified although in cases introducing additional fitting parameters can help with the situation but is well founded and understood, which limits the applicability of such approaches.

Recently, inspired by B. S. Shastry's \emph{extremely correlated Fermi liquid}\cite{shastry-2011-extrem-correl,shastry-2013-extrem-correl} (ECFL) which combined the Heisenberg- equations-of-motion (HEOM) and Schwinger-Dyson-equations-of-motion (SDEOM) to tackle the noncanonicality, W. Ding proposed an \emph{algebraic-dynamical theory}\cite{Ding-2022-algeb-dynam} (ADT) for strongly interacting many-body quantum Hamiltonians. ADT considers the complete set of noncanonical quantum algebras of the physical operators and the corresponding correlation functions, then using HEOM-SDEOM to establish a complete mapping from the state onto the state itself. This complete mapping, in principle, enable one to formulate a dynamical perturbation theory under an arbitrary perturbation when an exact eigenstate of the unperturbed Hamiltonian is given.

In this work, we study the exact ADT solutions of a set of two-spin problems for quantum spin-1/2. These two-spin models can serve as ``free theories'' for models with either fully dimerized ground states, such as a valence bond solid (VBS) state, or with states having only up to two-spin correlations, such as those of the 1D transverse field Ising models (TFIMs). With these free theories, we can construct dynamical perturbation theories to study more realistic models that can be constructed by perturbing these solvable models. As a proof of principle, we then perform dynamical perturbation calculations in the two-spin limit. The results are compared with exact diagonalization, showing quantitative agreement at the corresponding order of expansion in terms of the controlling parameter.

The rest of this article is organized as follows. We first review the ADT formalism briefly in Sec. \ref{sec-II}. In Sec. \ref{sec-III}, we study the two-spin limit of TFIMs. The classical Ising model is treated as the free theory, and the transverse field term is then applied as perturbation. Detailed ADT dynamical perturbation study is presented. In Sec. \ref{sec-VI}, study on the two-spin limit of Heisenberg-type models are discussed. Finally, we discuss the implications of our result

\section{The Algebraic-Dynamical-Theory Formalism for Quantum Spin-1/2}
\label{sec-II}
\subsection{the Algebraic-Dynamical-Theory Formalism}
\label{sec:orga12c55a}
Here we briefly introduce the ADT formalism. See Ref. \cite{Ding-2022-algeb-dynam} for a detailed discussion.
In ADT, in order to have a complete description of the dynamics of a quantum system, a \emph{complete  operator basis set} (COBS)\cite{Schwinger1960c,Fano1957}, which we denote as \(\mathscr{U} = \{ \hu^\alpha \}\) which satisfies the orthogonality and completeness conditions\cite{Fano1957}
\begin{align}
  \label{eq:1}
  \tr ((\hu^\alpha)^\dagger \hu^\beta) = C ~\delta_{\alpha \beta}, \\
  \hu^\alpha \hu^\beta = \sum_{\gamma} a^{\alpha \beta}_\gamma \hu^{\gamma},\label{eq:2}
\end{align}
where $C$ is a normalization factor which will be taken as 1 for convenience unless noted otherwise and $a_{\alpha \beta}^\gamma \in \mathbb{C}$.
Satisfying both the complete and the orthogonal conditions, each operator $\hO$ of $\mathscr{H}$ can be expanded into a sum of \(\hu^\alpha\)s,
\begin{align}
  \label{eq:3}
  \hO = \sum_\alpha \tr (\hO \hu^\alpha) \hu^\alpha .
\end{align}
It is convenient and conventional to decompose the algebras into a symmetric (bosonic) sector and an anti-symmetric (fermionic) sector
\begin{align}
  [\hu^\alpha, \hu^\beta ] = \sum_{\gamma} b^{\alpha \beta}_\gamma \hu^\gamma, \quad
  \{ \hu^\alpha, \hu^\beta \} = \sum_{l} f^{\alpha \beta}_\gamma \hu^\gamma, \label{eq:4}
\end{align}
with \(b^{\alpha \beta}_\gamma = a^{\alpha \beta}_\gamma - a^{\beta \alpha}_\gamma, \quad
  f^{\alpha \beta}_\gamma = a^{\alpha \beta}_\gamma + a^{\beta \alpha}_\gamma\).

Given a COBS $\mathscr{U}$, instead of specifying an arbitrary state as superposition of eigenstates or unit vectors of $\mathscr{H}$, the state can be specified by the expectation values of the bases operators $\ket{\{\ev{\hu^\alpha} \}}$ (as a vector), or simply as \( \ev{\mathscr{U}} \). In particular, according to \eqdisp{eq:4}, the density matrix, pure or mixed, can be written as \(\rho = \sum_\alpha \tr(\rho \hu^\alpha) \hu^\alpha = \sum_\alpha \ev{ \hu^\alpha} \hu^\alpha\).

Given a COBS $\mathscr{U}$ and a state \( \ev{\mathscr{U}} \), ADT further considers the complete set of (two-time) dynamical correlation functions (CSDCF) $\mathscr{G}$, whose elements $G^{\alpha \beta}_{\pm}[\i,\f]$'s are defined as
\begin{align}
  \label{eq:5}
  i G^{\alpha;\beta}_{\pm}[\i,\f] = \dev{\mT_\pm \Big( \hu^{\alpha} (t_i), \hu^{\beta} (t_f) \Big)},
\end{align}
where $\mT_\pm$ denotes time-ordering with the $\pm$ sign, $\dev{~}$ denotes fully dynamical correlations defined as
\begin{align}
  \label{eq:6}
    \dev{\hu^\alpha (t_i) \hu^\beta (t_f)} = \left\langle (\hu^{\alpha}(t_i)-\ev{\hu^\alpha}) (\hu^\beta(t_f) - \ev{\hu^\beta}) \right\rangle.
\end{align}
The equal-time-limit of $\mathscr{G}$ is a complete mapping $\ev{\mathscr{U} \times \mathscr{U}} \mapsto \ev{\mathscr{U}}$
\begin{align}
  i G^{\alpha; \beta}_{+}[\i,\i] = \ev{ \{ \hu^{\alpha} , \hu^{\beta} \}}/2 - \ev{\hu^{\alpha}} \ev{\hu^{\beta} }  ,\label{eq:7}\\
  i G^{\alpha; \beta}_{-}[\i,\i] = \ev{ [ \hu^{\alpha},\hu^{\beta} ]}/2, \label{eq:8}
\end{align}
as $\mathscr{G}$ covers the full algebra table and a convention of \(\theta(0)=1/2\) is adopted.

Next, we take a time-derivative on all the elements of CSDCF
and apply the HEOM to obtain a many-body analogue of Schwinger-Dyson-equations-of-motion (SDEOM):
\begin{align}
  \label{eq:9}
  \begin{split}
    i \pt_{t_i } G_{\pm}^{\alpha ;\beta}[\i,\f] & = \delta^{\alpha \beta}_{\pm} [\i,\f] + i \dev{\mT_\pm \big( [\hu^\alpha(t_i) , \mH], \hu^\beta(t_f) \big) }\\
    & = \delta^{\alpha \beta}_{\pm}[\i,\f] +  \sum_{\eta \gamma} h_\eta  b^{\alpha \eta}_\gamma G_{\pm}^{\gamma ;\beta}[\i,\f],
\end{split}
\end{align}
where $\delta^{\alpha \beta}_{\pm}[\i,\f] = 2 \delta(t_i - t_f)  i G^{\alpha; \beta}_{\mp}[\i,\i]$ and the Hamiltonian is given as $H= \sum_\alpha h^\alpha \hu^\alpha$.
Note that the equal-time-limit of SDEOM is also a complete mapping $i\pt_t \ev{\mathscr{U} \times \mathscr{U}} \mapsto \ev{\mathscr{U}}$.

For lattice problems, the COBS $\{\hu^{\alpha \beta \cdots}_{ij \cdots}\}$ is typically constructed from local COBS $\{\hu^\alpha_i\}$ as Cartesian products or their superposition $\hu^{\alpha \beta \dots}_{ij\dots} = \hu^\alpha_i \hu^\beta_j \cdots
$. When we represent a state with such many-body COBS, it is more physical to use their cumulant $\ev{\hu^\alpha_i \hu^\beta_j \dots}_c$ instead of the plain expectation values.

Finally, we can arrange the CSDCF into a vector form, and rewrite the complete set of SDEOM of the CSDCF in a matrix form:
\begin{align}
  \label{eq:10}
   i \pt_{t_i } [{\mathbf G}_{\pm}] = [\mathbf{\Delta}]_{\pm} [\i,\f] + [[\mathbf{L}]] \cdot [{\mathbf G}_{\pm}],
\end{align}
where we use $[\mathbf{\Delta}]_{\pm}$ as the vector form of $\{ \delta^{\alpha \beta}_{\pm} [\i,\f] \}$ (with ${\alpha}$ as the component index) and $ [[\mathbf{L}]] $ as the matrix form of $\{ \sum_{\eta} h_\eta  b^{\alpha \eta}_\gamma \}_{\alpha \gamma}$ (with different $ \alpha$ and $\gamma$ as the element indexes). For time-translational invariant problems, we can transform \eqdisp{eq:11} into frequency space as
\begin{align}
  \label{eq:11}
   \left( \omega I - [[\mathbf{L}]] \right) \cdot [{\mathbf G}_{\pm}] = [\mathbf{\Delta}]_{\pm} [\i,\f],
\end{align}
which can solved formally.
In this form, perturbation methods and theories of linear systems \cite{book-nayfeh-2008-pertu-methods,book-holmes-2012-intro-pertu} can be introduced.
The LHS of \eqdisp{eq:12} is determined by the Hamiltonian, disregarding the state of concern. Therefore, the RHS determines the pivoting for \eqdisp{eq:12} as a linear system. In other words, the nonzero terms of \([\mathbf{\Delta}]_{\pm} [\i,\f]\), i.e. the state. While exact solutions are generally difficult, if not impossible, it is reasonable to expect that a tractable of number of components could be sufficient to describe a system statistical-mechanically.

\subsection{Algebraic-Dynamical-Theory of a Single Spin-1/2}
\label{sec:orgb04c943}
For a single spin-1/2, COBS can be chosen as either \(\{\hat{I}, \hs^x, \hs^y, \hs^z \}\) or \(\{\hat{I}, \hs^+, \hs^-, \hs^z \}\). The latter is used more often for spin dynamics. We shall follow such convention, using the former to for wavefunctions or states and the latter for dynamics.
Its algebra table can be decomposed into a conventional form as:
\begin{align}
[\hs^\alpha, \hs^\beta] = i \epsilon^{\alpha \beta \gamma} \hs^\gamma,\quad \{\hs^+, \hs^-\} = 1, \{\hs^\pm, \hs^z\} = 0.\label{eq:12}
\end{align}
An arbitrary spin state, conventionally denoted by a wavefunction $\ket{\phi} = (a + i b) \ket{\ua} + (c + id) \ket{\da}$, is equivalently represent by COBS as $\{\ev{\hs^x} = a c+ bd, \ev{\hs^y} = ad-bc, \ev{\hs^z} = (a^2 + b^2 - c^2 -d^2)/2 \}$.
CSDCF of a single spin-1/2 $\{G^{\alpha ; \beta }_{\pm}[\i,\f]\}$ contains the usual single particle correlations studied in literature.

To compare with experiments, the dynamical structure factors (DSFs) are often required, which are the Fourier transform of the unordered two-time correlation functions \(S^{\alpha \beta}(\x,t) = \ev{\hs^\alpha_{\x}(t) \hs^\beta_{0} (0)}/(2\pi)\). We can express DSFs via \(\mathscr{G}\) as
\begin{align}
  \begin{split}
    & S^{\alpha \beta}(t) = -\frac{i}{4 \pi} (G^{\alpha\beta;R}_{+}(t) - G^{\alpha\beta;A}_{+}(t) \\
    & + G^{\alpha\beta;R}_{-}(t) - G^{\alpha\beta;A}_{-}(t)).
  \end{split}
  \label{eq:13}
\end{align}
With the above expression, the space-time DSF can be obtained by integrating the correlation functions in frequency space. For the rest of this work, we shall work with the causal Green's functions only, and simply inspect \( S^{\alpha \beta}[\omega] \sim -i (G^{\alpha\beta}_{+}[\omega] + G^{\alpha\beta}_{-}[\omega])/(4 \pi) \) without carrying out the subtraction \(G^R - G^A\).

\subsection{The Two-Spin Problem as Simplified Limits of Lattice Models}
\label{sec:org60f3f70}

Although the two-spin problems can be solved easily by diagonalization, the dynamical correlation functions have not been studied in detail. The two-spin problems can play the role of a Hubbard atom in Hubbard models, which is still being studied, generating useful insights\cite{thunstroem-2018-analy-inves} for lattice model studies.

For quantum spin models, this ``atom'' can describe states including i) a single bond correlation in RVB VBS etc, ii) ground states of TFIMs for which only two-spin cumulant correlations are present and iii) exactly solvable models with fully dimerized ground states such as the Shastry-Sutherland lattice model\cite{shastry-1981-exact-groun}, spin- Su-Schrieffer-Heeger (SSH) models\cite{su-1980-solit-excit-polyac,heeger-1988-solit-conduc-polym}, etc..
Therefore, detailed studies of of such atomic limits are necessary ground work for dynamical perturbation studies on top of these states and models. Moreover, analysis of the exact solutions can help us understand the analytic properties of the dynamical correlation functions in different states and contribute to improving the approximations one can make for similar states on a lattice.

\section{Two-spin limit of Transverse Field Ising Model}
\label{sec-III}
Consider an QSM Hamiltonian
\begin{align}
  \label{eq:14}
  \mH_{ab} = \mH_{0,ab} + \mH_{1,ab},
\end{align}
where \(\mH_{0,ab} = J_z \hs^z_a \hs^z_b - h^z (\hs^z_a + \hs^z_b)$, $\mH_{1,ab} = - h_x (\hs^x_a + \hs^x_b)\).
The Ising interaction term $J_z \hs^z_a \hs^z_b$ is considered as the ``free theory'', and since the $h^z$ term commutes with the Ising term. A transverse field term $\mH_1$ is taken as a perturbation. If we consider $J_z<0$, this toy model can be extended to a lattice as the transverse field Ising model when $h^z = 0$. When $J_z>0$ this model can be viewed as the atomic limit of the HM, as previously discussed while the $\mH_1$ can be interpreted as a Weiss mean field term stemming from the hopping. For the rest of this section, we only consider $J_z >0$ without losing generality.

\subsection{Grade-2 COBS}
\label{sec:org17b3d08}
To provide a ADT solutions, we first specify a grade-2 COBS for the problem. Since there are only two site involved, grade-2 is also the largest possible grade for the problem.
Consider two spin-1/2's $\hs_a$ and $\hs_b$.  We use the following grade-2 COBS construction:
\begin{align}
  \begin{split}
    &\hS^\alpha_{ab} = \hs_a^\alpha + \hs_b^\alpha,~
    \heta_{ab}^\alpha = \hs_a^\alpha - \hs_b^\alpha,\\
    & \hB_{A,ab}^{\gamma} =2 \varepsilon_{\alpha \beta \gamma} \hs_a^\alpha \hs_b^\beta ,\\
    & \hB_{S,ab}^{\gamma} =2 \varepsilon_{\alpha \beta \gamma}^2 \hs_a^\alpha \hs_b^\beta ,\\
   &  \hD_{ab}^\alpha = 2 \hs_a^\alpha \hs_b^\alpha,
  \end{split}
  \label{eq:15}
\end{align}
where the Einstein summation notation over dummy indices is assumed.

Although a direct product construction is equally valid and more convenient to implement in programming, this construction provides more physical insight which we shall discuss later. This particular construction follows from the geometric algebra\cite{Doran2009}. But for the discussion of dynamical correlation functions, we stick to the Cartesian COBS.
Since $B^\gamma_{S(A)ba} = \pm B^\gamma_{S(A)ab}$, we drop the $_{ab}$ index and denote $B_{S(A)}^\gamma = B^\gamma_{S(A)ab}$. So the grade-2 COBS is written as
\begin{align}
  \label{eq:16}
  \mathscr{U}_{g2} = \{\hI, \hS^\alpha, \heta^\alpha, \hB_S^\alpha, \hB_A^\alpha, \hD^\alpha \}.
\end{align}
Any state should be described by $\ev{\mathscr{U}_{g2}}$, which has 16 real numbers, exceeding the 8 real parameters allowed by considering the 4-dimensional $\hilbert_{ab}$. This reflects the algebraic constraints on the operators' expectation values.
While formally we choose $\mathscr{U}_{g2}$ as \eqdisp{eq:15} for convenience of algebras, we shall use $\mathscr{U}'_{g2} = \{\hI, \hs^\alpha_i, \hB_S^\alpha, \hB_A^\alpha, \hD^\alpha \}$, i.e. use local single particle operators $\hs^\alpha$ instead of the Fourier components, when discussing Green's functions.
\subsection{Exact Solutions of the States}
\label{sec:org7f9437a}
\subsubsection{Interacting ``Free Theory'': the Ising limit}
\label{sec:org683da1b}
The ground state solution of this model is the following:
\begin{equation}
\ket{\psi} =
\begin{cases}
  ~\ket{\da \da}, \quad \text{if $h_z <- J_z$}, \\
 ~\ket{\ua \ua}, \quad \text{if $h_z > J_z$},\\
 ~e^{i \frac{\phi}{2}}\cos\theta \ket{\ua \da} + e^{-i \frac{\phi}{2}} \sin \theta \ket{\da \ua},  \text{if $J_z > \abs{h_z}$}.
\end{cases}\label{eq:17}
\end{equation}
The interesting point to note is that for $J > \abs{h}$, the ground state can be arbitrary linear combination of the two states $\ket{\ua \da}$ and $\ket{\da \ua}$, reflecting a residual $U(1)$ symmetry. 
Alternatively, we use the ADT representation for the states or wavefunctions, i.e. by the expectation values of COBS.
Since our main focus is the last case, we give its ADT representation in Table \ref{tab:1}.

\begin{table}[H]
\centering
\begin{tabular}{|c|c|c|c|c|c|}
\hline
\(\ev{\hS^z}_c\) & 0 & \(\ev{\hS^x}_c\) & 0 & \(\ev{\hS^y}_c\) & 0\\
\hline
\(\ev{\heta^z}_c\) & \(\cos 2\theta\) & \(\ev{\heta^x}_c\) & 0 & \(\ev{\heta^y}_c\) & 0\\
\hline
\(\ev{\hB_S^z}_c\) & 0 & \(\ev{\hB_S^x}_c\) & 0 & \(\ev{\hB_S^y}_c\) & 0\\
\hline
\(\ev{\hB_A^z}_c\) & \(\sin(2\theta)\sin(\phi)\) & \(\ev{\hB_A^x}_c\) & 0 & \(\ev{\hB_A^y}_c\) & 0\\
\hline
\(\ev{\hD^z}_c\) & \(-\frac{\sin^2(2\theta)}{2}\) & \(\ev{\hD^x}_c\) & \(\frac{\sin(2\theta)\cos(\phi)}{2}\) & \(\ev{\hD^y}_c\) & \(\frac{\sin(2\theta ) \cos(\phi)}{2}\)\\
\hline
\end{tabular}
\caption{\label{tab:1}COBS presentation of the state \(e^{i \frac{\phi}{2}}\cos\theta \ket{\ua \da} + e^{-i \frac{\phi}{2}} \sin \theta \ket{\da \ua}\) with \(E = -J_z/4\).}

\end{table}
\subsubsection{Transverse Field as Perturbation}
\label{sec:org4f3b985}
When a sufficiently small transverse-field is applied, the \(U(1)\) symmetry reflected by the \(\phi\) angle is destroyed and a small transverse magnetization is produced. The exact results are given in Table \ref{tab:2}.
\begin{table}[H]
\centering
\begin{tabular}{|c|c|c|c|c|c|}
\hline
\(\ev{\hS^z}_c\) & 0 & \(\ev{\hS^x}_c\) & \(-\frac{4 h_x}{\sqrt{J_z^2+ 16 h_x^2}}\) & \(\ev{\hS^y}_c\) & 0\\
\hline
\(\ev{\heta^z}_c\) & 0 & \(\ev{\heta^x}_c\) & 0 & \(\ev{\heta^y}_c\) & 0\\
\hline
\(\ev{\hB_S^z}_c\) & 0 & \(\ev{\hB_S^x}_c\) & 0 & \(\ev{\hB_S^y}_c\) & 0\\
\hline
\(\ev{\hB_A^z}_c\) & 0 & \(\ev{\hB_A^x}_c\) & 0 & \(\ev{\hB_A^y}_c\) & 0\\
\hline
\(\ev{\hD^z}_c\) & \(-\frac{J_z/2}{\sqrt{J_z^2+ 16 h_x^2}}\) & \(\ev{\hD^x}_c\) & \(\frac{J_z^2/2}{J_z^2 + 16 h_x^2}\) & \(\ev{\hD^y}_c\) & \(\frac{J_z/2}{\sqrt{J_z^2+ 16 h_x^2}}\)\\
\hline
\end{tabular}
\caption{\label{tab:2}ADT presentation of the ground state for \(J_z > 4 h_x > 0\) with \(E=-\frac{\sqrt{16 h_x^2+J_z^2}}{4}\).}

\end{table}
\subsection{Green’s Functions of ``Free Theories''}
\label{sec:orgd65d9b7}
To accommodate \eqdisp{eq:17} and to simplify notations, we switch to notations of Green's functions in the form as \(G_{\pm}[\hO_1^\alpha;\hO_2^\beta](t \text{ or } \omega)\), where \(\hO^\alpha\) can be an element of \eqdisp{eq:17} or a product operator such as \(\hs^\alpha_i \hs^\beta_j\). We also omit \((t \text{ or } \omega)\) when there is no ambiguity.
Since it is too cumbersome to display the complete form of \eqdisp{eq:12}, we shall only put down and discuss necessary terms, i.e. the pivoting equations, for a given state.

Conventionally, the single particle Green's functions, i.e. \(G_{\pm}[\hs^\alpha_i ; \hs^\beta_j]\), can be solved from SDEOM, if and only the vertex functions due to interactions can be solved. For example, the typical single particle GFs of concern are those of \(\hs_i^\pm\). With only \(\mH_0\), the HEOM reads
\begin{align}
i\pt_t \hs^\alpha_{i} = \alpha (-J_z \hs^z_{\bar{i}} \hs^\alpha_{i} - h^z \hs^\alpha_{i}),\label{eq:18}
\end{align}
with $\alpha = \pm$, which leads to the following SDEOM
\begin{align}
  \begin{split}
  &  \alpha \omega G_{+}[\hs^\alpha_i;\hs^\beta_j] = \delta_{ij} \delta_{\alpha \beta } \ev{2 \hs^z_i} \\
  & - h^z G_{+}[\hs^\alpha_i;\hs^\beta_j] - J_z G_{+}[\hs^z_{\bar{i}} \hs^\alpha_i;\hs^\beta_j],
  \end{split}\label{eq:19}
\end{align}
for $\alpha,~ \beta,~ \gamma \in \{+,-\}$ and $\alpha$ in the equations is interpreted as a sign $\pm$.

Conventional approaches rely on the decoupling the vertex function \(\ev{\hs^z_{\bar{i}} \hs^\alpha_i;\hs^\beta_j} \rightarrow \ev{\hs^z_{\bar{i}}} \ev{\hs^\alpha_i;\hs^\beta_j}\), which is only exact for \emph{not-time-ordered correlation functions} of a product state. Otherwise, the decoupling requires extra free parameters, which is termed ``random phase approximation''(RPA) or Tyablikov-decoupling \cite{Kondo1972,tyablikov-1967-method-quant}. As we shall find next, for entangled states and time-ordered correlation functions, the vertex functions are independent dynamical degrees of freedom. RPA and its parameters are of a dynamical origin.

Consider a state specified in Table \ref{tab:1}, the single particle GFs and the associated vertex functions are
\begin{align}
  \begin{split}
    & G_{+}[\hs^\alpha_a;\hs^\beta_a] = \frac{\delta_{\alpha \bar{\beta}} \alpha (2 \ev{\hs^z_a} \omega - J_z/2) }{(\omega + J_z/2) (\omega - J_z/2)}, \\
    & G_{+}[\hs^\alpha_a;\hs^\beta_b] = \frac{ \delta_{\alpha \bar{\beta}} J_z  \ev{\{\hs_a^\alpha, \hs_b^\beta\}} }{(\omega + J_z/2) (\omega - J_z/2)}, \\
  & G_{+}[\hs^z_b \hs^\alpha_a;\hs^\beta_a] = \frac{\delta_{\alpha \bar{\beta}} (\ev{\hs^z_a} J_z - \omega)/2 }{(\omega + J_z/2) (\omega - J_z/2)},\\
   & G_{-}[\hs^\alpha_a;\hs^\beta_a] = \frac{\delta_{\alpha \bar{\beta}} (\omega - \alpha J_z \ev{\hs^z_i}) }{(\omega + J_z/2) (\omega - J_z/2)}, \\
   & G_{-}[\hs^\alpha_a;\hs^\beta_b] = \frac{ \delta_{\alpha \bar{\beta}} \ev{\{\hs_a^\alpha, \hs_b^\beta\}} \omega }{(\omega + J_z/2) (\omega - J_z/2)},\\
 & G_{-}[\hs^z_b \hs^\alpha_a;\hs^\beta_a] = \frac{\delta_{\alpha \bar{\beta}} \omega \ev{\hs^z_a} }{(\omega + J_z/2) (\omega - J_z/2)},\\
  \end{split}
  \label{eq:20}
\end{align}
where \(\bar{\pm} = \mp\).
First, we verify the solutions by taking the equal-time limit, which should reduce to \eqdisp{eq:8} and (\ref{eq:8}).
While the above expressions are valid for an arbitrary state of Table \ref{tab:1}, we shall discuss two of the most important ones: the fully polarized states and the fully entangled states.

\subsubsection{Product States}
\label{sec:org68a8a19}

For a polarized state, we verify some well-known physical properties to show the validity of the ADT solution. For a fully polarized state \(\ket{\phi} = \ket{\ua \da}\), the DSF is asymmetric between spin flip up and down processes. Indeed, we find \(S[\hs^+_a; \hs^-_a] = {1}{\omega + J_z/2}\) and \(S[\hs^-_a; \hs^+_a] = 0\). For an arbitrary state, we have \(S[\hs^\alpha_a; \hs^{\bar{\alpha}}_a] = {1/2 \pm \ev{\s^z_a}}{\omega + J_z/2}\).

We can also study the RPA for arbitrary states by taking the ratio between the single particle GFs and their corresponding vertex functions such as \(G[\hs^z_b \hs^+_a]; \hs^-_a]/G[\hs^+_a]; \hs^-_a] = (\omega - J_z \cos\theta /2 )/(J_z - 2 \omega \cos \theta)\). The ratio correctly reduces to \(- 1/2  = \ev{\hs^z_b}\) for \(\cos\theta = 1\), but generally it is a dynamical factor for entangled states.

\subsubsection{the Entangled States}
\label{sec:org91f197b}

For a fully entangled state in Table \ref{tab:1}, we have \(\cos (2\theta) = 0\). We also take \(\cos (\phi ) = 0\) so that the state of concern is either a triplet Bell state or a singlet: \(\ket{\psi} = (\ket{\ua \da} \pm \ket{\da \ua})/\sqrt{2}\). Then the only nonzero elements of \(\ev{\mathscr{U}_{g2}}\) are \ev{\hD^\alpha}. For the triplet, \(\ev{\hD^x}_c = \ev{\hD^y}_c = -\ev{\hD^z}_c = 1/2\). For the singlet, \(\ev{\hD^x}_c = \ev{\hD^y}_c = \ev{\hD^z}_c = -1/2\). In this case, SDEOM of single particles GFs are no longer pivotal since \(\ev{\hs_i^\alpha} =0\) (except for \(G_-[\hs_i^\pm, \hs_i^\mp]\)). As we shall find, it is necessary to study and use the many-body correlation functions as free theories for the perturbative calculation.
\subsection{Dynamical Perturbation Theory for Transverse Fields}
\label{sec-III-D}
\subsubsection{Simple 1st order perturbation}
\label{sec:org8d966dc}
As stated previously, we utilize the completeness of the algebras to compute variations of the state. When a transverse field is turned on, a uniform transverse magnetization \(\ev{\hs^x_a} = \ev{\hs^x_b} \propto h_x\) should be induced. Starting from the solutions for \(h_x =0\), we can make use of the following relations
\begin{align}
  \label{eq:21}
  & \ev{\hs^x_a} = i \ev{[\hs^z_a, \hs^y_a]}  = G_{-}[\hs^z_a;\hs^y_a](t=0).
\end{align}
Without $h_x$, $G_{-}^{(0)}[\hs^z_a;\hs^y_a] = 0$.

When $h_x$ is turned on, a new contribution $\propto h_x$ to HEOM $i \pt_t \hs^z_i  = i h_x \hs^y_i$. The corresponding new contribution to SDEOM reads
\begin{align}
i \pt_t G_{-}[\hs^z_a;\hs^y_a]  =  i h_x G_{-}[\hs^y_a;\hs^y_a].\label{eq:22}
\end{align}
Therefore, the leading order correction $G_{-}^{(1)}[\hs^z_a;\hs^y_a] = i h_x G_{-}^{(0)}[\hs^y_a;\hs^y_a]/\omega $. We find
\begin{align}
  \begin{split}
    \ev{\hs^x_a}_{1} \simeq i h_x \int d\omega  G_{-}^{(0)}[\hs^y_a;\hs^y_a]/\omega
    = \frac{ 4 h_x h^z \ev{\hs^z_a}_0}{(J_z)^2 - 4 (h^z)^2}  ,
  \end{split}\label{eq:23}
\end{align}
which agrees with diagonalization study well at sizable $h^z$ and $\ev{\hs^z}$ at the leading order of $h_x$.

However, this perturbative calculation fails as $h^z \rightarrow 0$ and $\ev{\hs^z} \rightarrow 0$. In contrast, the exact result is $ \ev{\hs^x_i} \simeq - h_x/J_z$ when $h^z =0$ \& $\ev{\hs^z} = 0$ and the exact ground state is almost a triplet state.
To correctly account for perturbation to $\ev{\hs^x_a}$ for $h^z =0$, we have to work with a many-body analogue of \eqdisp{eq:22}:
\begin{align}
  \label{eq:24}
  \begin{split}
    [\hs^z_a \hs^z_b, \hs^y_a \hs^z_b] = i \hs^x_a (\hs^z_b)^2
    \rightarrow \ev{\hs^x_a} = - 4 i \ev{[\hs^z_a \hs^z_b, \hs^y_a \hs^z_b]},
  \end{split}
\end{align}
where $(\hs^z)^2 = 1/4$ is applied.
Now we can utilize the following SDEOM (higher order terms in $h_x$ are ignored)
\begin{align}
  \label{eq:25}
  i \pt_{t} G_{-}[\hs^z_a \hs^z_b; \hs^y_a \hs^z_b] \simeq  - i h_x G_{-}[\hs^z_a \hs^y_b, \hs^y_a \hs^z_b]
\end{align}
to give a better perturbation for $\ev{\hs^x_a}$
\begin{align}
  \label{eq:26}
  \begin{split}
    \ev{\hs^x_a}_{1} 
    &\simeq \frac{- 4 h_x \ev{\hs^x_a \hs^x_b}_0}{J_z}.
  \end{split}
\end{align}
For the triplet state, $\ev{\hs^x_a \hs^x_b}_0 = 1/4$.
We must note that the exact numerical value is $-2 h_x/J_z$. A factor of 2 is missing.

While in both situations, results comparable to the exact ones up to certain accuracy can be obtained, we desire more systematic and controllable perturbation methods. The algebraic relations in Eq. \eqref{eq:21} and \eqref{eq:24} are exact. Therefore, we expect correct solutions through them to give consistent results, or even better, exactly the same results. Indeed, such calculations can be achieved by careful treatment of the corrections to vertex functions.

\subsubsection{Complete perturbation calculation}
\label{sec:org7b0b2c9}
To illustrate the validity of complete calculations, we use \eqdisp{eq:22} as an example.
We begin with the HEOM of \(\hs^y_i\) in the presence of the transverse field term, which reads
\begin{align}
  \begin{split}
    i\pt_t \hs^z_a =  i h_x \hs^y_a,\\
    i\pt_t \hs^y_a = - i h_x \hs^z_a + i J_z \hs^x_a \hs^z_b,\\
    i\pt_t (\hs^x_a \hs^z_b) = -i J_z \hs^y_a /4 + i h_x (\hs^x_a \hs^y_b),\\
    i\pt_t (\hs^x_a \hs^y_b) = - i h_x (\hs^x_a \hs^z_b).
  \end{split}\label{eq:27}
\end{align}
They lead to a set of modified SDEOM
\begin{align}
  \begin{split}
    &\omega G_{-}[\hs^z_a;\hs^y_a] = i h_x G_{-}[\hs^y_a;\hs^y_a],\\
    &\omega G_{-}[\hs^y_a;\hs^y_a] =  \frac{1}{2} - i h_x G_{-}[\hs^z_a;\hs^y_a] + i J_z G_{-}[\hs^x_a \hs^z_b;\hs^y_a],\\
    &\omega G_{-}[\hs^x_a \hs^z_b;\hs^y_a] = \frac{-i J_z}{4} G_{-}[\hs^y_a;\hs^y_a] + i h_x G_{-}[\hs^x_a \hs^y_b;\hs^y_a],\\
    &\omega G_{-}[\hs^x_a \hs^y_b;\hs^y_a] = -i h_x G_{-}[\hs^x_a \hs^z_b;\hs^y_a].
  \end{split}\label{eq:28}
\end{align}
Eq. \eqref{eq:28} is a closed set of equations, which can be diagonalized and gives
\begin{align}
  G_{-}[\hs^z_a;\hs^y_a] = -\frac{2 i h_x \left(h_x^2-\omega^2\right)}{4 h_x^4-8 h_x^2 \omega^2-J_z^2 \omega^2+4 \omega^4}. \label{eq:29}
\end{align}
Integrating over $\omega$, we now obtain the correct result for $\ev{\hs^x_a}_1 \simeq -2 h_x/J_z$.


We can complete the calculation through \eqdisp{eq:25} in the same way. The crucial difference is that, for \eqdisp{eq:25}, \([\Delta]_-\) contains relevant triplet correlations of the wavefunction, instead of being solely a constant. We can also identify additional algebraic relations that lead to similar calculations.
But in all cases, consistent perturbation for arbitrary states proves difficult. Different starting points mix different wavefunction components and energy spectral differently, so that the accuracy of simple perturbations starting from different algebraic relations varies, depending the state of concern.
\subsubsection{Unified Perturbation Formula via Variational Method}
\label{sec:org75e34fc}
In previous section, we find that different starting points give the same results but in different dynamical forms. Physically, we understand that the variation of states is just the consequence of minimization of the energy \(\ev{H_0 + H_1}\).
Therefore, we should expect a converged expression which accounts for variations in all components of the state in terms of the unperturbed state. For the above example, the exact induced $\ev{\hs^x_a}$ to the leading order in $h_x$ should receive contributions from all preexisting components of the unperturbed state as
\begin{align}
  \label{eq:30}
  \delta \ev{\hs^x_a} = h_x \left(d_z \ev{\hs^z_a}_0 + \sum_{\alpha} d_{\alpha\alpha} \ev{\hs^\alpha_a \hs^\alpha_b}_0 + \dots\right),
\end{align}
where $d_{\alpha\dots}$s are coefficients to be computed dynamically from the unperturbed state. Without further justification, we observe that $d_{\alpha\dots}$ is given by the simple perturbation starting from an algebraic relation giving \(\hs^x_a\).

In fact, such variational equations of states are known in classical Hamiltonian mechanics\cite{inbook-whittaker-1937-treatise}. A generic quantum mechanical version of such variational equations for arbitrary Hamiltonians and perturbation is also possible, which shall be presented elsewhere\cite{Ding-2022-varia-equat}.

\section{Two-spin limit of Heisenberg-type Models}
\label{sec-IV}
We discuss the Heisenberg-type models on the basis of the Ising limit with a in-plane interaction term:
\begin{align}
  \mathcal{H}_{\text{Heis.}} = \mathcal{H}_{0} + \mathcal{H}_{xy} = J_z \hs^z_a \hs^z_b + J_\perp( \hs^x_a \hs^x_b +  \hs^y_a \hs^y_b). \label{eq:31}
\end{align}
In the two-spin limit, \(H_{xy}\) lifts the degeneracy and make the ground state a spin singlet for \(J_z>0 ~\&~ J_\perp > 0\). In terms of COBS, it is specified by \(\ev{\hD^{\alpha}}_c = -1/2\). FM polarized states remain as excited eigenstates, but become the ground states for \(J_z < 0\). The perturbation to the singlet state in the two-spin limit is not interesting since it is full gapped and is rigid for infinitesimal perturbations when \(J_\perp\) is considered finite. The FM states form the ground-state manifold for \(J_z<0 ~\&~ J_\perp < 0\). More generally, we can consider arbitrary \(J_z ~\&~ J_\perp\) which corresponds to XXZ models on lattices. Therefore, we shall discuss both the singlet state and the FM states, without specifying the sign or relative amplitudes of \(J_z ~\&~ J_\perp\) and whether the state is the ground state or not.

\subsection{Green's Functions of ``Free Theories''}
\label{sec:orgb63ed1b}

\subsubsection{FM Product States}
\label{sec:org5b09a64}
Considering the state \(\ket{\phi} = \ket{\ua \ua}\) so that \(\ev{\hs^z_i} = 1/2\).  On a lattice, such fully polarized states are considered approximately describing a magnetically ordered state. Conventionally, excitations of a magnetic state are described by bosons as spin wave/magnon excitations. With ADT, we obtain the following bosonic single particle GFs:
\begin{align}
  \begin{split}
    & G_{+}[\hs^\alpha_a;\hs^\beta_a] = \frac{ \delta_{\alpha \bar{\beta}} ( \omega - \alpha J_z/2) }{(\omega - \alpha J_z/2)^2 - J_\perp^2/4}, \\
    & G_{+}[\hs^\alpha_a;\hs^\beta_b] = \frac{ - \delta_{\alpha \bar{\beta}} J_\perp/2}{( \omega - \alpha  J_z/2)^2 - J_\perp^2/4},
  \end{split}
  \label{eq:33}
\end{align}
for $\alpha,\beta = \pm$. It is instructive to examine the GFs in the "momentum space" at "\(k = 0 ~\&~ \pi \)" as \(G_+[\hs^\alpha;\hs^\beta]_{k=0 (k=\pi)} = ( G_{+}[\hs^\alpha_a;\hs^\beta_a] \pm G_{+}[\hs^\alpha_a;\hs^\beta_b] )/2 = \frac{2 \alpha}{\omega -(J_z \pm J_\perp)}\). Now we also set \(J_\perp = J_z = J\) in order to compare with traditional spin wave descriptions. We find that \(G_+[\hs^+;\hs^-]_{k=0} = (1/2) / \omega\) and \(G_+[\hs^+;\hs^-]_{k=\pi} = (1/2)/(\omega - J)\), which are consistent with a FM spin wave theory with a dispersion relation  \(\omega_k \propto J (1 - \cos k) \) due to a nearest-neighbor coupling in 1D. In fact, for Heisenberg models on lattices, RPA readily gives a spin-wave result.
Also similar as the TFIM case, the DSFs of a polarized state is zero for \( \hs^+\) excitations: \( S[\hs^-;\hs^+] = 0 \).

\subsubsection{the Singlet State}
\label{sec:orgc67ffe6}
Considering the single state \(\ket{\phi} = (\ket{\ua \da} - \ket{\da \ua})/\sqrt{2}\), which is the ground state for \(J_z~\&~ J_\perp > 0\). For simplicity, we take \(J_z = J_\perp = J\). The single GFs read
\begin{align}
 \label{eq:34}
  \begin{split}
    & G_{+}[\hs^\alpha_a;\hs^\beta_a] = \frac{ - 2 \delta_{\alpha \bar{\beta}} J }{\omega^2 - J^2},  ~ G_{+}[\hs^\alpha_a;\hs^\beta_b] = \frac{2 \delta_{\alpha \bar{\beta}} J }{\omega^2 - J^2},\\
    & G_{-}[\hs^\alpha_a;\hs^\beta_a] = \frac{ 2 \delta_{\alpha \bar{\beta}} \omega }{\omega^2 - J^2},
     G_{-}[\hs^\alpha_a;\hs^\beta_b] = \frac{ - 2 \delta_{\alpha \bar{\beta}} \omega }{\omega^2 - J^2},
  \end{split}
\end{align}
Similar as the TFIM case, RPA breaks down for entangled states. These single particle GFs are all fully gapped out, in contrast to those of FM states.
However, there still are man-body correlation functions with low excitation energies. For example, we find
\begin{align}
\label{eq:35}
  \begin{split}
  G_-[\hs^+_a\hs^-_b;\hs^-_a\hs^+_b] = \frac{J^2}{ \omega (\omega^2 - J^2)}, \\
  G_-[\hs^z_a\hs^z_b;\hs^-_a\hs^+_b] = \frac{1}{2 \omega}.
  \end{split}
\end{align}

\subsection{Dynamical Perturbations}
\label{sec:org6ac7c47}
Similar as the TFIM case, ADT formalism allow us to perform dynamical perturbative calculations for arbitrary small perturbations to the Hamiltonian. Since the main purpose of this work is for proof of principle, which is mostly fulfilled by discussion of Sec. \ref{sec-III-D}, we shall only discuss dynamical perturbations for the two ``free theories'' of the previous section briefly.

For a FM product state, it is similar to the product state of two-spin Ising model. For the isotropic case with \(J_z = J_\perp\) and in the absence of external fields, the FM states are 3-fold degenerate, forming the \(S=1\) subspace. An external field perturbation along any direction singles out the polarized state. If we consider a field along the \(\hat{z}-\)direction, it is straightforward to see since the zeroth order contribution \(\ev{\delta H}_0 = h_x \ev{\sum_i \hs^z_i}_0\) to the energy already gives the right result. However, if we take a field along \(\hat{x}-\)direction, we need to consider a parameterized superposition of these three states and optimize the energy to determine the parameters, similar as how we solve the TFIM perturbation with the a 2-fold degeneracy.

For the singlet state, it is the unique ground state for \(J > 0\), the single particle excitations, as shown by the poles of single particle GFs, are all gapped out. We expect that no infinitesimal perturbations should change the state immediately. However, the results of the previous section can serve as the ``free theory'' for lattice models with a fully dimerized ground state.
\section{Implications for Lattice Models}
\label{sec-V}
\subsection{Transverse Field Ising on Lattices}
\label{sec:org88bf8a0}
For standard 1D TFIM with only nearest-neighbor Ising interaction, the ground state across the phase diagram is completely specified by up to the second cumulants: \(\ket{\Psi} = \{\ev{\hs^\alpha_i}, \ev{\hs^\alpha_i \hs^\alpha_j}_c\}\). Despite it is exactly solvable by the Jordan-Wigner transformation, computation of some components of the dynamical correlation functions is still difficult.
ADT provides a new route to obtain the dynamical correlations since both the states and the operator algebra are solvable. However, ADT can go beyond the existing exact results and study perturbations that spoil the integrability.

For TFIM beyond the standard form, such as in higher dimensional lattices with or without frustration, DAT allows perturbation studies of quantum models starting from the classical limit where the states or static correlation functions can be obtained by exact solutions, diagrammatic methods for classical Ising models\cite{Skryabin1988} and numerical methods.

\subsection{Heisenberg-type Models}
\label{sec:orgc432663}
The two-spin singlet state is the basic building block of valence bond solid (VBS) states. A large class of solvable models possess a ground state which is a product state of isolated, full singlets, such as a  spin-Peierls state in 1D\cite{pytte-1974-peier-instab}, the Shastry-Sutherland lattice model\cite{shastry-1981-exact-groun}, etc.. Results of this work enable us to investigate perturbation theories to such states analytically.

Furthermore, the two-spin limit results are potentially useful in describing more difficult problems such as a uniform RVB state and the coexistence of AFM and RVB singlet correlation.
For example, an uniform RVB state at mean field level can be specified by \(\ev{\hD^\alpha_{<ij>}}_c = \text{const}\) for nearest-neighbor bonds. However, \(\ev{\hD^\alpha_{<ij>}}_c\) can not be maximized to the value of a full singlet of two spins. Additional constraining conditions need to be applied, which we leave to future works.

\section{Conclusion}
\label{sec-VI}
In this work, we apply the recently proposed \emph{algebraic dynamical theory} to quantum spin-1/2 models in a two-spin limit. We use ADT to obtain the exact solutions of dynamical correlation functions in different parameter setting for different states of relevant lattice models of interests. We verify the validity of ADT solutions by comparing with exact diagonalization calculations. We further perform a dynamical perturbation calculation for a two-spin transverse field Ising model, treating the transverse fields as perturbation. We obtain quantitatively accurate results for the variation of the state at \(\mathcal{O}(h_x/J_z))\). We also discuss solutions of Heisenberg-type models. For a fully polarized state, we verify the validity of the ``random phase approximation''(RPA) or Tyablikov-decoupling \cite{Kondo1972,tyablikov-1967-method-quant} in this case, and reproduce a spin-wave spectrum for the single particle correlation functions. However, when the state is not a pure product state, RPA starts to breakdown and the ratio between the vertex function and the corresponding single particle GF become a dynamical factor. Our results may also help to improve RPA in such situations. Finally, we discuss the implications of these results for relevant lattice models.
\section*{Acknowledgements}
We thank R. Yu and J. D. Wu for many beneficial discussions, F. C. Zhang for encouragement of this work. The work at Anhui University was supported by the Anhui Provincial Natural Science Foundation Young Scientist Grant number 1908085QA35 and the Startup Grant number S020118002/002 of Anhui University. WD thanks support from Kavli Institute for Theoretical Sciences for visits during the work.
\appendix
\section{Grade-2 Algebra Table (fermionic algebras)}
\label{sec:orgc77c042}
The new set of algebra are computed and given below as
The commutation algebras:
\begin{widetext}
  \begin{align}
    \begin{split}
    [S^\alpha,~S^\beta] = i \varepsilon_{\alpha \beta \gamma} S^\gamma,\quad
    [\eta^\alpha,~\eta^\beta] = i \varepsilon_{\alpha \beta \gamma} S^\gamma, \quad
    [B_S^\alpha, ~B_S^\beta] = - i \varepsilon_{\alpha \beta \gamma} S^\gamma,\quad
    [B_A^\alpha, ~B_A^\beta] = + i \varepsilon_{\alpha \beta \gamma} S^\gamma,\\
    [S^\alpha,~\eta^\beta] = i \varepsilon_{\alpha \beta \gamma} \eta^\gamma,\quad  [S^\alpha, ~B_S^\beta] = - i \varepsilon_{\alpha \beta \gamma} B_S^\gamma + 2 i \delta_{\alpha \beta}\varepsilon_{ij\alpha} D^j,\quad
    [S^\alpha, ~B_A^\beta] = i \varepsilon_{\alpha \beta \gamma} B_A^\gamma\\
    [\eta^\alpha,~B_S^\beta] = + i \varepsilon^2_{\alpha \beta \gamma} B_A^\gamma,\quad
    [\eta^\alpha, B_A^\beta]  = -i \varepsilon^2_{\alpha \beta \gamma} B_S^\gamma + 2 i \delta_{\alpha \beta}\varepsilon^2_{ij\alpha} D^j,\quad
    [B_S^\alpha, ~B_A^\beta] = i \varepsilon_{\alpha \beta \gamma}^2 \eta^\gamma,\\
    [D^\alpha, S^\beta] = i  (1 - \delta_{\alpha \beta})  \sgn[\alpha-\beta] (-1)^{\beta-\alpha+1} B_S^\beta,\quad
    [D^\alpha, B_{S}^\beta] = - i  (1 - \delta_{\alpha \beta})  \sgn[\alpha-\beta] (-1)^{\alpha-\beta} S^\beta,\\
   [D^\alpha,~D^\beta]= 0,  \quad  [D^\alpha, \eta^\beta] = i   (1 - \delta_{\alpha \beta})  B_A^\beta ,\quad
    [D^\alpha, B_{A}^\beta] = -i   (1 - \delta_{\alpha \beta})  \eta^\beta,
    \end{split}
    \label{eq:a1}
  \end{align}
  The anti-commutation algebras:
  \begin{align}
    \begin{split}
    \{\hS^\alpha, \hS^\beta\} = \delta_{\alpha \beta} (1 + 2 \hD^\alpha) +\epsilon^2_{\alpha \beta \gamma} \hB_S^\gamma,
    \quad \{\heta^\alpha, \heta^\beta\} = \delta_{\alpha \beta} (1 - 2 \hD^\alpha) - \epsilon^2_{\alpha \beta \gamma} \hB_S^\gamma,\\
    \{\hB_S^\alpha,\hB_S^\beta\} = \delta_{\alpha \beta} (1 + 2 \hD^\alpha) - \epsilon^2_{\alpha \beta \gamma} \hB_S^\gamma,\quad
    \{\hB_A^\alpha,\hB_A^\beta\} = \delta_{\alpha \beta} (1 - 2 \hD^\alpha) - \epsilon^2_{\alpha \beta \gamma} \hB_S^\gamma,\
    \\
     \{\hS^\alpha,\heta^\beta\} = - \epsilon_{\alpha \beta \gamma} \hB_A^\gamma,\quad
     \{\hS^\alpha,\hB_S^\beta\} = \epsilon^2_{\alpha \beta \gamma} \hS^\gamma,\quad
     \{\hS^\alpha,\hB_A^\beta\} = \epsilon_{\alpha \beta \gamma} \heta^\gamma,\quad
     \{\hS^\alpha,\hD^\beta\} = \delta_{\alpha \beta} \hS^\alpha,\\
     \{\heta^{\alpha}, \hB_S^{\beta}\} = - \epsilon^2_{\alpha \beta \gamma} \heta^\gamma, \quad
     \{\heta^{\alpha}, \hB_A^{\beta}\} = - \epsilon_{\alpha \beta \gamma} \hS^\gamma,\quad
     \{\heta^{\alpha}, \hD^{\beta}\} = - \delta_{\alpha \beta } \heta^\alpha,
     \\
     \{\hB_S^{\alpha}, \hB_A^{\beta}\} = - \epsilon^2_{\alpha \beta \gamma} \hB_S^\gamma, \quad
     \{\hB_S^{\alpha}, \hD^{\beta}\} = \delta_{\alpha \beta } \hB_S^\alpha,\quad
     \{\hB_A^{\alpha}, \hD^{\beta}\} = - \delta_{\alpha \beta } \hB_A^\alpha,\quad
    \{D^\alpha,~D^\beta\}  =   \delta_{\alpha \beta}/2 - \varepsilon_{\alpha \beta \gamma}^2 D^\gamma,
    \end{split}
    \label{eq:a2}
  \end{align}
\end{widetext}

\bibliography{../../library,../../../org/notes,../../../org/research/refs}

\begin{thebibliography}{37}%
\makeatletter
\providecommand \@ifxundefined [1]{%
 \@ifx{#1\undefined}
}%
\providecommand \@ifnum [1]{%
 \ifnum #1\expandafter \@firstoftwo
 \else \expandafter \@secondoftwo
 \fi
}%
\providecommand \@ifx [1]{%
 \ifx #1\expandafter \@firstoftwo
 \else \expandafter \@secondoftwo
 \fi
}%
\providecommand \natexlab [1]{#1}%
\providecommand \enquote  [1]{``#1''}%
\providecommand \bibnamefont  [1]{#1}%
\providecommand \bibfnamefont [1]{#1}%
\providecommand \citenamefont [1]{#1}%
\providecommand \href@noop [0]{\@secondoftwo}%
\providecommand \href [0]{\begingroup \@sanitize@url \@href}%
\providecommand \@href[1]{\@@startlink{#1}\@@href}%
\providecommand \@@href[1]{\endgroup#1\@@endlink}%
\providecommand \@sanitize@url [0]{\catcode `\\12\catcode `\$12\catcode
  `\&12\catcode `\#12\catcode `\^12\catcode `\_12\catcode `\%12\relax}%
\providecommand \@@startlink[1]{}%
\providecommand \@@endlink[0]{}%
\providecommand \url  [0]{\begingroup\@sanitize@url \@url }%
\providecommand \@url [1]{\endgroup\@href {#1}{\urlprefix }}%
\providecommand \urlprefix  [0]{URL }%
\providecommand \Eprint [0]{\href }%
\providecommand \doibase [0]{https://doi.org/}%
\providecommand \selectlanguage [0]{\@gobble}%
\providecommand \bibinfo  [0]{\@secondoftwo}%
\providecommand \bibfield  [0]{\@secondoftwo}%
\providecommand \translation [1]{[#1]}%
\providecommand \BibitemOpen [0]{}%
\providecommand \bibitemStop [0]{}%
\providecommand \bibitemNoStop [0]{.\EOS\space}%
\providecommand \EOS [0]{\spacefactor3000\relax}%
\providecommand \BibitemShut  [1]{\csname bibitem#1\endcsname}%
\let\auto@bib@innerbib\@empty
\bibitem [{\citenamefont {Ising}(1925)}]{Ising1925}%
  \BibitemOpen
  \bibfield  {author} {\bibinfo {author} {\bibfnamefont {E.}~\bibnamefont
  {Ising}},\ }\bibfield  {title} {\bibinfo {title} {{Beitrag zur Theorie des
  Ferromagnetismus}},\ }\href {https://doi.org/10.1007/BF02980577} {\bibfield
  {journal} {\bibinfo  {journal} {Zeitschrift f{\"{u}}r Phys.}\ }\textbf
  {\bibinfo {volume} {31}},\ \bibinfo {pages} {253} (\bibinfo {year}
  {1925})}\BibitemShut {NoStop}%
\bibitem [{\citenamefont {Bethe}(1931)}]{bethe-1931-zur-theor-metal}%
  \BibitemOpen
  \bibfield  {author} {\bibinfo {author} {\bibfnamefont {H.}~\bibnamefont
  {Bethe}},\ }\bibfield  {title} {\bibinfo {title} {Zur theorie der metalle},\
  }\href {https://doi.org/10.1007/bf01341708} {\bibfield  {journal} {\bibinfo
  {journal} {Zeitschrift f\"{u}r Physik}\ }\textbf {\bibinfo {volume} {71}},\
  \bibinfo {pages} {205} (\bibinfo {year} {1931})}\BibitemShut {NoStop}%
\bibitem [{\citenamefont {Haldane}(1983)}]{haldane-1983-nonlin-field}%
  \BibitemOpen
  \bibfield  {author} {\bibinfo {author} {\bibfnamefont {F.~D.~M.}\
  \bibnamefont {Haldane}},\ }\bibfield  {title} {\bibinfo {title} {Nonlinear
  field theory of large-spin heisenberg antiferromagnets: Semiclassically
  quantized solitons of the one-dimensional easy-axis n{\'e}el state},\ }\href
  {https://doi.org/10.1103/physrevlett.50.1153} {\bibfield  {journal} {\bibinfo
   {journal} {Phys. Rev. Lett.}\ }\textbf {\bibinfo {volume} {50}},\ \bibinfo
  {pages} {1153} (\bibinfo {year} {1983})}\BibitemShut {NoStop}%
\bibitem [{\citenamefont {Affleck}\ \emph {et~al.}(1987)\citenamefont
  {Affleck}, \citenamefont {Kennedy}, \citenamefont {Lieb},\ and\ \citenamefont
  {Tasaki}}]{affleck-1987-rigor-resul}%
  \BibitemOpen
  \bibfield  {author} {\bibinfo {author} {\bibfnamefont {I.}~\bibnamefont
  {Affleck}}, \bibinfo {author} {\bibfnamefont {T.}~\bibnamefont {Kennedy}},
  \bibinfo {author} {\bibfnamefont {E.~H.}\ \bibnamefont {Lieb}},\ and\
  \bibinfo {author} {\bibfnamefont {H.}~\bibnamefont {Tasaki}},\ }\bibfield
  {title} {\bibinfo {title} {Rigorous results on valence-bond ground states in
  antiferromagnets},\ }\href {https://doi.org/10.1103/physrevlett.59.799}
  {\bibfield  {journal} {\bibinfo  {journal} {Phys. Rev. Lett.}\ }\textbf
  {\bibinfo {volume} {59}},\ \bibinfo {pages} {799} (\bibinfo {year}
  {1987})}\BibitemShut {NoStop}%
\bibitem [{\citenamefont {i.~Tomonaga}(1950)}]{tomonaga-1950-remar-bloch}%
  \BibitemOpen
  \bibfield  {author} {\bibinfo {author} {\bibfnamefont {S.}~\bibnamefont
  {i.~Tomonaga}},\ }\bibfield  {title} {\bibinfo {title} {Remarks on bloch's
  method of sound waves applied to many-fermion problems},\ }\href
  {https://doi.org/10.1143/ptp/5.4.544} {\bibfield  {journal} {\bibinfo
  {journal} {Prog. Theor. Phys.}\ }\textbf {\bibinfo {volume} {5}},\ \bibinfo
  {pages} {544} (\bibinfo {year} {1950})}\BibitemShut {NoStop}%
\bibitem [{\citenamefont {Luttinger}(1963)}]{luttinger-1963-exact-solub}%
  \BibitemOpen
  \bibfield  {author} {\bibinfo {author} {\bibfnamefont {J.~M.}\ \bibnamefont
  {Luttinger}},\ }\bibfield  {title} {\bibinfo {title} {An exactly soluble
  model of a many‐fermion system},\ }\href
  {https://doi.org/10.1063/1.1704046} {\bibfield  {journal} {\bibinfo
  {journal} {J. Math. Phys.}\ }\textbf {\bibinfo {volume} {4}},\ \bibinfo
  {pages} {1154} (\bibinfo {year} {1963})}\BibitemShut {NoStop}%
\bibitem [{\citenamefont {HALDANE}(1994)}]{haldane-1994-demon-of}%
  \BibitemOpen
  \bibfield  {author} {\bibinfo {author} {\bibfnamefont {F.~D.~M.}\
  \bibnamefont {HALDANE}},\ }\bibinfo {title} {Demonstration of the "luttinger
  liquid" character of bethe-ansatz-soluble models of 1-d quantum fluids},\ in\
  \href {https://doi.org/10.1142/9789812798268_0036} {\emph {\bibinfo
  {booktitle} {Exactly Solvable Models of Strongly Correlated Electrons}}},\
  \bibinfo {series and number} {Exactly Solvable Models of Strongly Correlated
  Electrons}\ (\bibinfo  {publisher} {WORLD SCIENTIFIC},\ \bibinfo {year}
  {1994})\ pp.\ \bibinfo {pages} {441--443}\BibitemShut {NoStop}%
\bibitem [{\citenamefont {Savary}\ and\ \citenamefont
  {Balents}(2017)}]{Savary2017}%
  \BibitemOpen
  \bibfield  {author} {\bibinfo {author} {\bibfnamefont {L.}~\bibnamefont
  {Savary}}\ and\ \bibinfo {author} {\bibfnamefont {L.}~\bibnamefont
  {Balents}},\ }\bibfield  {title} {\bibinfo {title} {{Quantum spin liquids: a
  review}},\ }\href {https://doi.org/10.1088/0034-4885/80/1/016502} {\bibfield
  {journal} {\bibinfo  {journal} {Reports Prog. Phys.}\ }\textbf {\bibinfo
  {volume} {80}},\ \bibinfo {pages} {016502} (\bibinfo {year}
  {2017})}\BibitemShut {NoStop}%
\bibitem [{\citenamefont {Zhou}\ \emph {et~al.}(2017)\citenamefont {Zhou},
  \citenamefont {Kanoda},\ and\ \citenamefont {Ng}}]{Zhou2017}%
  \BibitemOpen
  \bibfield  {author} {\bibinfo {author} {\bibfnamefont {Y.}~\bibnamefont
  {Zhou}}, \bibinfo {author} {\bibfnamefont {K.}~\bibnamefont {Kanoda}},\ and\
  \bibinfo {author} {\bibfnamefont {T.-K.}\ \bibnamefont {Ng}},\ }\bibfield
  {title} {\bibinfo {title} {{Quantum spin liquid states}},\ }\href
  {https://doi.org/10.1103/RevModPhys.89.025003} {\bibfield  {journal}
  {\bibinfo  {journal} {Rev. Mod. Phys.}\ }\textbf {\bibinfo {volume} {89}},\
  \bibinfo {pages} {025003} (\bibinfo {year} {2017})}\BibitemShut {NoStop}%
\bibitem [{\citenamefont {Kitaev}(2006)}]{kitaev-2006-anyon-exact}%
  \BibitemOpen
  \bibfield  {author} {\bibinfo {author} {\bibfnamefont {A.}~\bibnamefont
  {Kitaev}},\ }\bibfield  {title} {\bibinfo {title} {Anyons in an exactly
  solved model and beyond},\ }\href {https://doi.org/10.1016/j.aop.2005.10.005}
  {\bibfield  {journal} {\bibinfo  {journal} {Ann. Physics}\ }\textbf {\bibinfo
  {volume} {321}},\ \bibinfo {pages} {2} (\bibinfo {year} {2006})}\BibitemShut
  {NoStop}%
\bibitem [{\citenamefont {Dyson}(1956)}]{Dyson1956}%
  \BibitemOpen
  \bibfield  {author} {\bibinfo {author} {\bibfnamefont {F.~J.}\ \bibnamefont
  {Dyson}},\ }\bibfield  {title} {\bibinfo {title} {{General Theory of
  Spin-Wave Interactions}},\ }\href {https://doi.org/10.1103/PhysRev.102.1217}
  {\bibfield  {journal} {\bibinfo  {journal} {Phys. Rev.}\ }\textbf {\bibinfo
  {volume} {102}},\ \bibinfo {pages} {1217} (\bibinfo {year}
  {1956})}\BibitemShut {NoStop}%
\bibitem [{\citenamefont {{S. V. Maleev}}(1958)}]{Maleev1958}%
  \BibitemOpen
  \bibfield  {author} {\bibinfo {author} {\bibnamefont {{S. V. Maleev}}},\
  }\bibfield  {title} {\bibinfo {title} {{Scattering of slow neutrons in
  ferromagnets}},\ }\href@noop {} {\bibfield  {journal} {\bibinfo  {journal}
  {Sov. Phys. JETP}\ }\textbf {\bibinfo {volume} {6}},\ \bibinfo {pages} {776}
  (\bibinfo {year} {1958})}\BibitemShut {NoStop}%
\bibitem [{\citenamefont {Auerbach}\ and\ \citenamefont
  {Arovas}(2008)}]{auerbach-2008-schwinger-boson}%
  \BibitemOpen
  \bibfield  {author} {\bibinfo {author} {\bibfnamefont {A.}~\bibnamefont
  {Auerbach}}\ and\ \bibinfo {author} {\bibfnamefont {D.~P.}\ \bibnamefont
  {Arovas}},\ }\href@noop {} {\bibinfo {title} {Schwinger bosons approaches to
  quantum antiferromagnetism}} (\bibinfo {year} {2008}),\ \Eprint
  {https://arxiv.org/abs/0809.4836} {arXiv:0809.4836 [cond-mat.str-el]}
  \BibitemShut {NoStop}%
\bibitem [{\citenamefont {Abrikosov}(1965)}]{abrikosov-1965-elect-scatt}%
  \BibitemOpen
  \bibfield  {author} {\bibinfo {author} {\bibfnamefont {A.~A.}\ \bibnamefont
  {Abrikosov}},\ }\bibfield  {title} {\bibinfo {title} {Electron scattering on
  magnetic impurities in metals and anomalous resistivity effects},\ }\href
  {https://doi.org/10.1103/physicsphysiquefizika.2.5} {\bibfield  {journal}
  {\bibinfo  {journal} {Phys. Phys. Fiz.}\ }\textbf {\bibinfo {volume} {2}},\
  \bibinfo {pages} {5} (\bibinfo {year} {1965})}\BibitemShut {NoStop}%
\bibitem [{\citenamefont {Affleck}\ and\ \citenamefont
  {Marston}(1988)}]{affleck-1988-large-nlimit}%
  \BibitemOpen
  \bibfield  {author} {\bibinfo {author} {\bibfnamefont {I.}~\bibnamefont
  {Affleck}}\ and\ \bibinfo {author} {\bibfnamefont {J.~B.}\ \bibnamefont
  {Marston}},\ }\bibfield  {title} {\bibinfo {title} {{Large-$N$ limit of the
  Heisenberg-Hubbard Model: Implications for High-T$_c$ superconductors}},\
  }\href {https://doi.org/10.1103/physrevb.37.3774} {\bibfield  {journal}
  {\bibinfo  {journal} {Phys. Rev. B}\ }\textbf {\bibinfo {volume} {37}},\
  \bibinfo {pages} {3774} (\bibinfo {year} {1988})}\BibitemShut {NoStop}%
\bibitem [{\citenamefont {Holstein}(1940)}]{Holstein1940}%
  \BibitemOpen
  \bibfield  {author} {\bibinfo {author} {\bibfnamefont {T.}~\bibnamefont
  {Holstein}},\ }\bibfield  {title} {\bibinfo {title} {{Field Dependence of the
  Intrinsic Domain Magnetization of a Ferromagnet}},\ }\href
  {https://doi.org/10.1103/PhysRev.58.1098} {\bibfield  {journal} {\bibinfo
  {journal} {Phys. Rev.}\ }\textbf {\bibinfo {volume} {58}},\ \bibinfo {pages}
  {1098} (\bibinfo {year} {1940})}\BibitemShut {NoStop}%
\bibitem [{\citenamefont {Lu}\ \emph {et~al.}(2017)\citenamefont {Lu},
  \citenamefont {Cho},\ and\ \citenamefont
  {Vishwanath}}]{lu-2017-unific-boson}%
  \BibitemOpen
  \bibfield  {author} {\bibinfo {author} {\bibfnamefont {Y.-M.}\ \bibnamefont
  {Lu}}, \bibinfo {author} {\bibfnamefont {G.~Y.}\ \bibnamefont {Cho}},\ and\
  \bibinfo {author} {\bibfnamefont {A.}~\bibnamefont {Vishwanath}},\ }\bibfield
   {title} {\bibinfo {title} {Unification of bosonic and fermionic theories of
  spin liquids on the kagome lattice},\ }\href
  {https://doi.org/10.1103/physrevb.96.205150} {\bibfield  {journal} {\bibinfo
  {journal} {Phys. Rev. B}\ }\textbf {\bibinfo {volume} {96}},\ \bibinfo
  {pages} {205150} (\bibinfo {year} {2017})}\BibitemShut {NoStop}%
\bibitem [{\citenamefont {Vaks}\ \emph {et~al.}(1968)\citenamefont {Vaks},
  \citenamefont {Larkin},\ and\ \citenamefont {Pikin}}]{Vaks-1968-spin-waves}%
  \BibitemOpen
  \bibfield  {author} {\bibinfo {author} {\bibfnamefont {V.~G.}\ \bibnamefont
  {Vaks}}, \bibinfo {author} {\bibfnamefont {A.~I.}\ \bibnamefont {Larkin}},\
  and\ \bibinfo {author} {\bibfnamefont {S.~A.}\ \bibnamefont {Pikin}},\
  }\bibfield  {title} {\bibinfo {title} {Spin waves and correlation functions
  in a ferromagnetic},\ }\href@noop {} {\bibfield  {journal} {\bibinfo
  {journal} {Sov. Phys. Jetp}\ }\textbf {\bibinfo {volume} {26}},\ \bibinfo
  {pages} {647} (\bibinfo {year} {1968})}\BibitemShut {NoStop}%
\bibitem [{\citenamefont {Kondo}\ and\ \citenamefont
  {Yamaji}(1972)}]{Kondo1972}%
  \BibitemOpen
  \bibfield  {author} {\bibinfo {author} {\bibfnamefont {J.}~\bibnamefont
  {Kondo}}\ and\ \bibinfo {author} {\bibfnamefont {K.}~\bibnamefont {Yamaji}},\
  }\bibfield  {title} {\bibinfo {title} {{Green's-Function Formalism of the
  One-Dimensional Heisenberg Spin System}},\ }\href
  {https://doi.org/10.1143/PTP.47.807} {\bibfield  {journal} {\bibinfo
  {journal} {Prog. Theor. Phys.}\ }\textbf {\bibinfo {volume} {47}},\ \bibinfo
  {pages} {807} (\bibinfo {year} {1972})}\BibitemShut {NoStop}%
\bibitem [{\citenamefont {Tyablikov}(1967)}]{tyablikov-1967-method-quant}%
  \BibitemOpen
  \bibfield  {author} {\bibinfo {author} {\bibfnamefont {S.~V.}\ \bibnamefont
  {Tyablikov}},\ }\href {https://doi.org/10.1007/978-1-4899-7182-1} {\emph
  {\bibinfo {title} {Methods in the Quantum Theory of Magnetism}}}\ (\bibinfo
  {publisher} {Springer US},\ \bibinfo {year} {1967})\BibitemShut {NoStop}%
\bibitem [{\citenamefont {FROBRICH}\ and\ \citenamefont
  {KUNTZ}(2006)}]{frobrich-2006-many-body}%
  \BibitemOpen
  \bibfield  {author} {\bibinfo {author} {\bibfnamefont {P.}~\bibnamefont
  {FROBRICH}}\ and\ \bibinfo {author} {\bibfnamefont {P.}~\bibnamefont
  {KUNTZ}},\ }\bibfield  {title} {\bibinfo {title} {Many-body green's function
  theory of heisenberg films},\ }\href
  {https://doi.org/10.1016/j.physrep.2006.07.002} {\bibfield  {journal}
  {\bibinfo  {journal} {Phys. Rep.}\ }\textbf {\bibinfo {volume} {432}},\
  \bibinfo {pages} {223} (\bibinfo {year} {2006})}\BibitemShut {NoStop}%
\bibitem [{\citenamefont {Shastry}(2011)}]{shastry-2011-extrem-correl}%
  \BibitemOpen
  \bibfield  {author} {\bibinfo {author} {\bibfnamefont {B.~S.}\ \bibnamefont
  {Shastry}},\ }\bibfield  {title} {\bibinfo {title} {Extremely correlated
  fermi liquids},\ }\href {https://doi.org/10.1103/physrevlett.107.056403}
  {\bibfield  {journal} {\bibinfo  {journal} {Phys. Rev. Lett.}\ }\textbf
  {\bibinfo {volume} {107}},\ \bibinfo {pages} {056403} (\bibinfo {year}
  {2011})}\BibitemShut {NoStop}%
\bibitem [{\citenamefont {Shastry}(2013)}]{shastry-2013-extrem-correl}%
  \BibitemOpen
  \bibfield  {author} {\bibinfo {author} {\bibfnamefont {B.~S.}\ \bibnamefont
  {Shastry}},\ }\bibfield  {title} {\bibinfo {title} {Extremely correlated
  fermi liquids: the formalism},\ }\href
  {https://doi.org/10.1103/physrevb.87.125124} {\bibfield  {journal} {\bibinfo
  {journal} {Phys. Rev. B}\ }\textbf {\bibinfo {volume} {87}},\ \bibinfo
  {pages} {125124} (\bibinfo {year} {2013})}\BibitemShut {NoStop}%
\bibitem [{\citenamefont {Ding}(2022{\natexlab{a}})}]{Ding-2022-algeb-dynam}%
  \BibitemOpen
  \bibfield  {author} {\bibinfo {author} {\bibfnamefont {W.}~\bibnamefont
  {Ding}},\ }\bibfield  {title} {\bibinfo {title} {Algebraic-dynamical theory
  for quantum many-body hamiltonians: A formalized approach to strongly
  interacting systems}\ }\href {https://doi.org/10.48550/ARXIV.2202.12082}
  {10.48550/ARXIV.2202.12082} (\bibinfo {year}
  {2022}{\natexlab{a}})\BibitemShut {NoStop}%
\bibitem [{\citenamefont {Schwinger}(1960)}]{Schwinger1960c}%
  \BibitemOpen
  \bibfield  {author} {\bibinfo {author} {\bibfnamefont {J.}~\bibnamefont
  {Schwinger}},\ }\bibfield  {title} {\bibinfo {title} {{Unitary Operator
  Bases}},\ }\href {https://doi.org/10.1073/pnas.46.4.570} {\bibfield
  {journal} {\bibinfo  {journal} {Proc. Natl. Acad. Sci.}\ }\textbf {\bibinfo
  {volume} {46}},\ \bibinfo {pages} {570} (\bibinfo {year} {1960})}\BibitemShut
  {NoStop}%
\bibitem [{\citenamefont {Fano}(1957)}]{Fano1957}%
  \BibitemOpen
  \bibfield  {author} {\bibinfo {author} {\bibfnamefont {U.}~\bibnamefont
  {Fano}},\ }\bibfield  {title} {\bibinfo {title} {{Description of States in
  Quantum Mechanics by Density Matrix and Operator Techniques}},\ }\href
  {https://doi.org/10.1103/RevModPhys.29.74} {\bibfield  {journal} {\bibinfo
  {journal} {Rev. Mod. Phys.}\ }\textbf {\bibinfo {volume} {29}},\ \bibinfo
  {pages} {74} (\bibinfo {year} {1957})}\BibitemShut {NoStop}%
\bibitem [{\citenamefont {Nayfeh}(2008)}]{book-nayfeh-2008-pertu-methods}%
  \BibitemOpen
  \bibfield  {author} {\bibinfo {author} {\bibfnamefont {A.~H.}\ \bibnamefont
  {Nayfeh}},\ }\href@noop {} {\emph {\bibinfo {title} {Perturbation methods}}}\
  (\bibinfo  {publisher} {John Wiley \& Sons},\ \bibinfo {year}
  {2008})\BibitemShut {NoStop}%
\bibitem [{\citenamefont {Holmes}(2012)}]{book-holmes-2012-intro-pertu}%
  \BibitemOpen
  \bibfield  {author} {\bibinfo {author} {\bibfnamefont {M.~H.}\ \bibnamefont
  {Holmes}},\ }\href@noop {} {\emph {\bibinfo {title} {Introduction to
  perturbation methods}}},\ Vol.~\bibinfo {volume} {20}\ (\bibinfo  {publisher}
  {Springer Science \& Business Media},\ \bibinfo {year} {2012})\BibitemShut
  {NoStop}%
\bibitem [{\citenamefont {Thunstr{\"o}m}\ \emph {et~al.}(2018)\citenamefont
  {Thunstr{\"o}m}, \citenamefont {Gunnarsson}, \citenamefont {Ciuchi},\ and\
  \citenamefont {Rohringer}}]{thunstroem-2018-analy-inves}%
  \BibitemOpen
  \bibfield  {author} {\bibinfo {author} {\bibfnamefont {P.}~\bibnamefont
  {Thunstr{\"o}m}}, \bibinfo {author} {\bibfnamefont {O.}~\bibnamefont
  {Gunnarsson}}, \bibinfo {author} {\bibfnamefont {S.}~\bibnamefont {Ciuchi}},\
  and\ \bibinfo {author} {\bibfnamefont {G.}~\bibnamefont {Rohringer}},\
  }\bibfield  {title} {\bibinfo {title} {Analytical investigation of
  singularities in two-particle irreducible vertex functions of the hubbard
  atom},\ }\href {https://doi.org/10.1103/physrevb.98.235107} {\bibfield
  {journal} {\bibinfo  {journal} {Phys. Rev. B}\ }\textbf {\bibinfo {volume}
  {98}},\ \bibinfo {pages} {235107} (\bibinfo {year} {2018})}\BibitemShut
  {NoStop}%
\bibitem [{\citenamefont {Shastry}\ and\ \citenamefont
  {Sutherland}(1981)}]{shastry-1981-exact-groun}%
  \BibitemOpen
  \bibfield  {author} {\bibinfo {author} {\bibfnamefont {B.~S.}\ \bibnamefont
  {Shastry}}\ and\ \bibinfo {author} {\bibfnamefont {B.}~\bibnamefont
  {Sutherland}},\ }\bibfield  {title} {\bibinfo {title} {Exact ground state of
  a quantum mechanical antiferromagnet},\ }\href
  {https://doi.org/10.1016/0378-4363(81)90838-x} {\bibfield  {journal}
  {\bibinfo  {journal} {Physica B+C}\ }\textbf {\bibinfo {volume} {108}},\
  \bibinfo {pages} {1069} (\bibinfo {year} {1981})}\BibitemShut {NoStop}%
\bibitem [{\citenamefont {Su}\ \emph {et~al.}(1980)\citenamefont {Su},
  \citenamefont {Schrieffer},\ and\ \citenamefont
  {Heeger}}]{su-1980-solit-excit-polyac}%
  \BibitemOpen
  \bibfield  {author} {\bibinfo {author} {\bibfnamefont {W.~P.}\ \bibnamefont
  {Su}}, \bibinfo {author} {\bibfnamefont {J.~R.}\ \bibnamefont {Schrieffer}},\
  and\ \bibinfo {author} {\bibfnamefont {A.~J.}\ \bibnamefont {Heeger}},\
  }\bibfield  {title} {\bibinfo {title} {Soliton excitations in
  polyacetylene},\ }\href {https://doi.org/10.1103/physrevb.22.2099} {\bibfield
   {journal} {\bibinfo  {journal} {Phys. Rev. B}\ }\textbf {\bibinfo {volume}
  {22}},\ \bibinfo {pages} {2099} (\bibinfo {year} {1980})}\BibitemShut
  {NoStop}%
\bibitem [{\citenamefont {Heeger}\ \emph {et~al.}(1988)\citenamefont {Heeger},
  \citenamefont {Kivelson}, \citenamefont {Schrieffer},\ and\ \citenamefont
  {Su}}]{heeger-1988-solit-conduc-polym}%
  \BibitemOpen
  \bibfield  {author} {\bibinfo {author} {\bibfnamefont {A.~J.}\ \bibnamefont
  {Heeger}}, \bibinfo {author} {\bibfnamefont {S.}~\bibnamefont {Kivelson}},
  \bibinfo {author} {\bibfnamefont {J.~R.}\ \bibnamefont {Schrieffer}},\ and\
  \bibinfo {author} {\bibfnamefont {W.~P.}\ \bibnamefont {Su}},\ }\bibfield
  {title} {\bibinfo {title} {Solitons in conducting polymers},\ }\href
  {https://doi.org/10.1103/revmodphys.60.781} {\bibfield  {journal} {\bibinfo
  {journal} {Rev. Modern Phys.}\ }\textbf {\bibinfo {volume} {60}},\ \bibinfo
  {pages} {781} (\bibinfo {year} {1988})}\BibitemShut {NoStop}%
\bibitem [{\citenamefont {Doran}\ and\ \citenamefont
  {Lasenby}(2003)}]{Doran2009}%
  \BibitemOpen
  \bibfield  {author} {\bibinfo {author} {\bibfnamefont {C.}~\bibnamefont
  {Doran}}\ and\ \bibinfo {author} {\bibfnamefont {A.}~\bibnamefont
  {Lasenby}},\ }\href {https://doi.org/10.1017/CBO9780511807497} {\emph
  {\bibinfo {title} {{Geometric Algebra for Physicists}}}}\ (\bibinfo
  {publisher} {Cambridge University Press},\ \bibinfo {year}
  {2003})\BibitemShut {NoStop}%
\bibitem [{\citenamefont {Whittaker}(1937)}]{inbook-whittaker-1937-treatise}%
  \BibitemOpen
  \bibfield  {author} {\bibinfo {author} {\bibfnamefont {E.}~\bibnamefont
  {Whittaker}},\ }\bibinfo {title} {A treatise on the analytical dynamics of
  particles and rigid bodies}\ (\bibinfo  {publisher} {Cambridge University
  Press},\ \bibinfo {year} {1937})\ Chap.\ \bibinfo {chapter} {X,
  112}\BibitemShut {NoStop}%
\bibitem [{\citenamefont {Ding}(2022{\natexlab{b}})}]{Ding-2022-varia-equat}%
  \BibitemOpen
  \bibfield  {author} {\bibinfo {author} {\bibfnamefont {W.}~\bibnamefont
  {Ding}},\ }\bibfield  {title} {\bibinfo {title} {Algebraic-dynamical
  variational equations of states for quantum many-body hamiltonians},\
  }\href@noop {} {\bibfield  {journal} {\bibinfo  {journal} {in preparation.}\
  } (\bibinfo {year} {2022}{\natexlab{b}})}\BibitemShut {NoStop}%
\bibitem [{\citenamefont {Izyumov}\ and\ \citenamefont
  {Skryabin}(1988)}]{Skryabin1988}%
  \BibitemOpen
  \bibfield  {author} {\bibinfo {author} {\bibfnamefont {Y.~A.}\ \bibnamefont
  {Izyumov}}\ and\ \bibinfo {author} {\bibfnamefont {Y.~N.}\ \bibnamefont
  {Skryabin}},\ }\href@noop {} {\emph {\bibinfo {title} {{Statistical Mechanics
  of Magnetically Ordered Systems}}}}\ (\bibinfo  {publisher} {Consultants
  Bureau},\ \bibinfo {address} {New York},\ \bibinfo {year} {1988})\BibitemShut
  {NoStop}%
\bibitem [{\citenamefont {Pytte}(1974)}]{pytte-1974-peier-instab}%
  \BibitemOpen
  \bibfield  {author} {\bibinfo {author} {\bibfnamefont {E.}~\bibnamefont
  {Pytte}},\ }\bibfield  {title} {\bibinfo {title} {Peierls instability in
  heisenberg chains},\ }\href {https://doi.org/10.1103/physrevb.10.4637}
  {\bibfield  {journal} {\bibinfo  {journal} {Phys. Rev. B}\ }\textbf {\bibinfo
  {volume} {10}},\ \bibinfo {pages} {4637} (\bibinfo {year}
  {1974})}\BibitemShut {NoStop}%
\end{thebibliography}%
\end{document}